\documentclass{amsproc}

\usepackage[parfill]{parskip}    
\usepackage{amsrefs, amsmath}
\usepackage{amsfonts, mathrsfs}
\usepackage{amssymb}
\usepackage{amscd}
\usepackage{cleveref}
\usepackage{fullpage, color}
\usepackage[labelformat=empty]{subfig}
\usepackage{booktabs}
\usepackage{xypic}
\usepackage[dvips]{graphicx,epsfig}
\usepackage{enumerate}

\BibSpec{article}{%
+{}{\sc \PrintAuthors} {author}
+{,}{ \textrm} {title}
+{,}{ \textit} {journal}
+{,}{ } {volume}
+{}{ \IfEmptyBibField{journal}{}{\PrintDate{date}}} {transition}
+{,}{ } {pages}
+{,}{ } {status}
+{.}{} {transition}
+{}{ Preprint\IfEmptyBibField{date}{}{ \PrintDate{date}}, available at \tt} {eprint}
+{.}{} {transition}
}

\BibSpec{book}{%
+{}{\sc \PrintAuthors} {author}
+{,}{ \textit} {title}
+{.}{ } {series}
+{,}{ } {volume}
+{.}{ } {publisher}
+{,}{ } {place}
+{,}{ } {date}
+{.}{} {transition}
}

\BibSpec{collection.article}{
+{} {\sc \PrintAuthors} {author}
+{,} { \textrm } {title}
+{,} { in \it \PrintConference} {conference}
+{,} { pp.~} {pages}
+{.} {\PrintBook} {book}
+{,} { \PrintDateB} {date}
+{.} {} {transition}
}

\BibSpec{innerbook}{
+{.} { \emph} {title}
+{.} { } {part}
+{:} { \emph} {subtitle}
+{.} { } {series}
+{,} { } {volume}
+{.} { Edited by \PrintNameList} {editor}
+{.} { Translated by \PrintNameList}{translator}
+{.} { \PrintContributions} {contribution}
+{.} { } {publisher}
+{.} { } {organization}
+{,} { } {address}
+{,} { \PrintEdition} {edition}
+{,} { \PrintDateB} {date}
+{.} { } {note}
}

\allowdisplaybreaks[3]

\crefname{equation}{Eq.}{Eqs.}
\crefname{eqnarray}{Eq.}{Eqs.}
\crefname{conj}{Conjecture}{Conjectures}
\crefname{lem}{Lemma}{Lemmas}
\crefname{thm}{Theorem}{Theorems}
\crefname{rmk}{Remark}{Remarks}
\crefname{prop}{Proposition}{Propositions}
\crefname{section}{Section}{Sections}
\crefname{appendix}{Appendix}{Appendices}
\crefname{cor}{Corollary}{Corollaries}
\crefname{figure}{Figure}{Figures}
\crefname{example}{Example}{Examples}

\DeclareMathOperator{\tr}{Tr}

\newcommand{\de}{{\partial}}

\newcommand{\rd}{\mathrm{d}}
\newcommand{\ri}{\mathrm{i}}
\newcommand{\re}{\mathrm{e}}

\newcommand{\bbZ}{\mathbb{Z}}
\newcommand{\bbR}{\mathbb{R}}
\newcommand{\bbC}{\mathbb{C}}
\newcommand{\bbP}{\mathbb{P}}

\def\bary{\begin{array}} 
\def\eary{\end{array}} 
\def\ben{\begin{enumerate}} 
\def\een{\end{enumerate}}
\def\bit{\begin{itemize}} 
\def\eit{\end{itemize}}
\def\nn{\nonumber} 

\newcommand{\cO}{\mathcal{O}}
\newcommand{\cT}{\mathcal{T}}

\newcommand{\LL}{\mathcal{L}}

\newcommand{\cK}{\mathcal{K}}
\newcommand{\cN}{\mathcal{N}}
\newcommand{\cW}{\mathcal{W}}
\newcommand{\cG}{\mathcal{G}}
\newcommand{\cA}{\mathcal{A}}

\newcommand{\cB}{\mathcal{B}}
\newcommand{\cF}{\mathcal{F}}

\newcommand{\cM}{\mathcal M}

\def\L{\Lambda}

\def\beq{\begin{equation}}                     %
\def\eeq{\end{equation}}                       %
\def\bea{\begin{eqnarray}}                     
\def\eea{\end{eqnarray}}
\def\bary{\begin{array}} 
\def\eary{\end{array}} 
\def\ben{\begin{enumerate}} 
\def\een{\end{enumerate}}
\def\bit{\begin{itemize}} 
\def\eit{\end{itemize}}
\def\nn{\nonumber} 
\def\de {\partial}

\theoremstyle{plain}

\newtheorem{thm}{Theorem}[section]

\newtheorem*{conj*}{Conjecture}

\newtheorem*{cor*}{Corollary}

\newtheorem{defn}{Definition}[section]

\theoremstyle{definition}

\newtheorem{example}{Example}[section]

\newcommand{\GIT}[1]{/\!\!/_{\kern-.2em #1 \kern0.1em}}

\renewcommand{\l}{\left}
\renewcommand{\r}{\right}

\def\bred{\begin{color}{red}}
\def\ered{\end{color}}

\def\bes{\begin{subequations}}
\def\ees{\end{subequations}}

\begin{document}

\title{On the Gopakumar--Ooguri--Vafa correspondence for Clifford--Klein 3-manifolds}

\author{Andrea Brini}
\address{Department of Mathematics,
Imperial College London, 180 Queen's Gate, SW7 2AZ, London, United Kingdom
}
\address{On leave from IMAG, Univ. Montpellier, CNRS, Montpellier, France}
\email{a.brini@imperial.ac.uk}

\begin{abstract}

Gopakumar, Ooguri and Vafa famously proposed the existence of a correspondence
between a topological gauge theory on one hand -- $\mathrm{U}(N)$ Chern--Simons theory on the
three-sphere -- and a topological string theory on the other -- the topological
A-model on the resolved conifold. On the physics side, this
duality provides a concrete instance of the large $N$ gauge/string
correspondence where exact computations can be performed in detail;
mathematically, it puts forward a triangle of striking relations between quantum invariants (Reshetikhin--Turaev--Witten) of knots and 3-manifolds, curve-counting invariants (Gromov--Witten/Donaldson--Thomas)
of local Calabi-Yau 3-folds, and the Eynard--Orantin recursion for a specific
class of spectral curves. I here survey recent results on the most general frame of validity of this
correspondence and discuss some of its implications.

\end{abstract}

\maketitle
\tableofcontents

\section{Introduction}

In a series of works \cite{Gopakumar:1998ki, Ooguri:1999bv, Ooguri:2002gx},
 Gopakumar, Ooguri and Vafa postulated and gave strong evidence for an identity
 between two rather different physical theories: on the one hand, ${\rm U}(N)$ Chern--Simons theory on $S^3$
 \cite{Witten:1988hf}, and on the other, topologically A-twisted string theory
 on the resolved conifold, $\mathrm{Tot}(\cO^{\oplus 2}_{\bbP^1}(-1))$. From
 the physics side, such
 identification is a concrete example of 't Hooft's idea \cite{'tHooft:1973jz}  that perturbative
 gauge theories with classical gauge groups in the $1/N$ expansion should be
 equivalent to some first-quantised string theory on a given background. The
 topological nature of both Chern--Simons theory and the A-model topological
 string allow for detailed checks of the correspondence, which might be
 regarded as a simplified setting for gauge/string dualities for the type II
 superstring, such as the AdS/CFT correspondence \cite{Maldacena:1997re}.

Mathematically, the implications of the Gopakumar--Ooguri--Vafa (GOV)
correspondence are perhaps even more noteworthy: the correspondence ties
together, in a highly non-trivial way, two different
theories of geometric invariants having {\it a priori} little resemblance to
each other. In Witten's landmark paper \cite{Witten:1988hf}
Chern--Simons observables were proposed to give rise
to topological invariants of framed 3-manifolds and links therein in light of the
Schwarz-type topological invariance of the quantum theory.
The relation to rational conformal field theory leads in particular to the skein relations
typical of knot invariants such as the HOMFLY-PT and Kaufmann
invariants, and more generally, to the Reshetikhin--Turaev invariants arising
in the representation theory of quantum groups and modular tensor categories 
\cite{Reshetikhin:1990pr,Reshetikhin:1991tc}.
The A-type topological string side, in turn, also enjoys a mathematical definition -- of a
quite different flavour. The partition function for a given
Calabi--Yau target $X$ is a formal generating function of various virtual counts of curves in $X$, either via
stable maps \cite{Kontsevich:1994na} or ideal sheaves \cite{MR1634503}. In
particular, the GOV correspondence asserts that the asymptotic expansion of
the Reshetikhin--Turaev invariant associated to the quantum group
$U_{q}(\mathrm{sl}_N)$ at large $N$ and fixed $q^N$ equates the formal Gromov--Witten
potential of $X$ in the genus expansion. This has a B-model counterpart due to
recent developments in higher genus toric mirror symmetry a
\cite{Bouchard:2007ys,Eynard:2012nj}, where the same genus expansion can be phrased in terms of
the Eynard--Orantin topological recursion \cite{Eynard:2007kz} on the
Hori--Iqbal--Vafa mirror curve of $X$ \cite{Hori:2000kt,Hori:2000ck}.

The GOV correspondence has had a profound impact for both communities
involved. In Gromov--Witten / Donaldson--Thomas theory, it has laid the foundations of the use of large $N$ dualities to
solve the topological string on toric backgrounds \cite{Aganagic:2002qg,
  Aganagic:2003db, moop} and to obtain all-genus results that went well beyond
the existing localisation computations at the time, as well as
some striking results for the intersection theory on moduli spaces of
curves \cite{Marino:2001re} and, via the relation of Chern--Simons theory to
random matrices, an embryo of the remodeling proposal
\cite{Marino:2002fk,Marino:2006hs,Bouchard:2007ys}. In the other direction, the
integral structure of BPS invariants leads to non-trivial
constraints for the structure of quantum knot invariants
\cite{Labastida:2000yw,Marino:2009mw, Liu:2007kv}.

Since the original correspondence of \cite{Gopakumar:1998ki, Ooguri:1999bv} was confined to the case where the gauge
group is the unitary group $\mathrm{U}(N)$, the base manifold is $S^3$, and the knot is
the trivial knot, a natural question to ask is
whether a similar connection could be generalised to other classical
gauge groups \cite{Sinha:2000ap, Bouchard:2004iu,Bouchard:2004ri}, knots other
than the unknot \cite{Labastida:2000zp, Labastida:2000yw, Brini:2011wi} (see
also \cite{Aganagic:2013jpa} for a significant generalisation, in a rather
different setting), as well as categorified/refined invariants of various
types \cite{Gukov:2004hz,Aganagic:2011sg}. A further natural direction would
be to seek the broadest generalisation of the correspondence in its original form beyond the
basic case of the three-sphere; this would require to provide a description of the string
dual of Chern--Simons both in terms of Gromov--Witten theory and of the
Eynard--Orantin theory on a specific spectral curve setup. This program was
initiated in \cite{Aganagic:2002wv} (see also
\cite{Halmagyi:2003ze,Halmagyi:2003mm,iolpq}) for the case of lens spaces, and
what is presumably its widest frame of validity has been recently described in
\cite{Borot:2015fxa,Brini:2017gfi}. 

This short survey is meant as an overview of results contained in
\cite{Borot:2015fxa,Brini:2017gfi}, putting them into context and highlighting their
implications for quantum invariants and Calabi--Yau enumerative geometry
alike. It is structured as follows: we first give a lightning review of the
three-way relation between matrix models, intersection theory on $\overline{\cM}_{g,n}$,
and the Eynard--Orantin recursion. We then move on to review the
uplift of these notions to the GOV correspondence for $S^3$. Finally, we discuss its
generalisation to the setting of Clifford--Klein 3-manifolds and outline the
proof of the B-model side of the correspondence for the case of the
L\^e--Murakami--Ohtsuki invariant \cite{MR1604883}.

\subsection*{Acknowledgements} I would like to thank my collaborators
Ga\"etan Borot, Bertrand Eynard, Luca Griguolo, Marcos Mari\~no, Domenico
Seminara and Alessandro Tanzini for sharing their insights and educating me
on various aspects of the Chern--Simons/topological strings correspondence as
well as for most enjoyable collaborations on this topic. This paper is a
write-up of material presented at a talk at the 2016 AMS Von Neumann Symposium on {\it Topological
  Recursion and its Influence in Analysis, Geometry, and Topology}, July 4–8,
2016, at Charlotte (NC); I wish to express my thanks to Motohico Mulase and
Chiu-Chu Melissa Liu for the wonderful scientific environment at the workshop
and for giving me the opportunity to present my work. This research was
partially supported by a fellowship under the ERC-2015-CoG ``GWT''.

\section{A 0-dimensional aper\c cu: matrix models, the topological recursion and enumerative geometry}

As a warm-up for the description of the triangle of relations linking 3D TQFT, curve-counting invariants and
the topological recursion, let us consider the much more basic setting of 0D
QFTs -- namely, matrix models. For any $N,r\in\bbZ^+$, let $\mathrm{U}(N)$ denote the rank-$N$ unitary group, $H_N$ be its Lie algebra of
hermitian matrices, and for $x$ in a formal neighbourhood of the origin of
$\bbC^r$ we let $\rd \mu_N(x)$ be a family of formally finite Ad-invariant measures, absolutely continuous
with respect to the standard gaussian measure $\rd \mu^{(0)}_N=\rd \mu_N(0)$ on $H_N\simeq
\bbR^{N^2}$. The typical setup here is
\beq
\rd \mu_N(x) = \cN_N \prod_{i=1}^N \rd M_{ii}\prod_{i<j}^N \rd \mathfrak{Re}M_{ij}
\rd \mathfrak{Im}M_{ij} \re^{-\frac{N}{2t} \mathrm{Tr} M^2 + \sum_{j}x_j
  \mathrm{Tr} M^j}=\cN_N \rd\mu^{(0)}_N \re^{\sum_{j}x_j \mathrm{Tr} M^j}
\label{eq:mun}
\eeq
where $\cN_N \in \bbC^\star$, and the exponential deformation in the last equality
should be treated as formal. For fixed $t$, we will denote by $Z_N(x)$ the formal total mass of
$\rd\mu_N(x)$,
\beq
Z_N(x)=\int_{H_N} \rd\mu_N(x) \in \bbC[[x]],
\label{eq:zn}
\eeq
and for any collection of positive integers $k_1, \dots, k_h$, $h\geq 1$, we write 
\beq
\mathsf{W}_{k_1, \dots, k_h;N}(x) \triangleq \de_{x_{k_1},\dots, x_{k_h}}^{\sum_i k_i}\log
Z_{N}(x) \in \bbC[[x]]
\label{eq:w}
\eeq
for the cumulants of the measure. \cref{eq:mun,eq:zn} define respectively
a {\it formal hermitian 1-matrix model} and its partition function, and we are
particularly interested in their formal behaviour as $N\to \infty$. It is a
matter of book-keeping in the application of Wick's theorem to
\cref{eq:zn,eq:w} to show that both $\ln Z_N$ and
$W_{k_1, \dots, k_m}$ have a formal connected ribbon graph ({\it fat-graph}) expansion
\cite{'tHooft:1973jz, Brezin:1977sv}
\beq
\ln Z_N(x)=  \sum_{g \geq 0} N^{2-2g} \cF_g(x), 
\label{eq:1/N}
\eeq
\beq
\mathsf{W}_{k_1, \dots, k_h;N}(x) = \sum_{g \geq 0}N^{2-2g-h}\cW_{g,h; k_1,
  \dots, k_h}(x).
\label{eq:1/Nw}
\eeq
The $1/N$ expansion of \cref{eq:1/N,eq:1/Nw} lends itself to two interpretations. 
\begin{itemize}
\item The
first one stems from the observation that, as the dual of a ribbon graph is a polygonulation of a connected, oriented
topological 2-manifold, this $1/N$ expansion can be regarded as a genus
expansion in a sum over random polygonulations of a Riemann surface. The $1/N$
expansion gives then an answer to an {\it enumeration problem}: it is easy to show in
particular, when $\cN_N=(t/N)^{N(N+1)/2} (\pi/N)^{-N^2/2} 2^{-N}$, that the
coefficients of the formal expansion in $x$ have an enumerative meaning as a
count of topologically inequivalent dual graphs. One consequence  is that, as
(metric) ribbon graphs can be used as a means to describe simplicial
decompositions of the moduli space $\cM_{g,n}$ of pointed Riemann surfaces via
the Strebel correspondence, the enumerative content of \cref{eq:1/N} provides a combinatorial answer
to topological questions on this moduli space \cite{MR963064, MR918455, Distler:1990mt}.
\item The second (physical) interpretation
comes from a special case of the observation, due to 't Hooft, that the perturbative $1/N$
expansion of a $\mathrm{U}(N)$ gauge theory with only adjoint fields takes the
shape of the expansion in the string coupling of a first quantised closed string
theory, possibly with probe branes insertions \cite{'tHooft:1973jz}; in a
sense the observation above that the enumeration problem of fatgraphs is
related to intersection-theoretic problems on moduli spaces of curves is the most basic
example of this phenomenon. Indeed, 1-matrix
models are possibly the most basic avatars of such QFTs, and it is only
natural to ask {\it what is the string dual of the formal 1-matrix model?}
This is
a sharp, and already non-trivial question in the context of {\it gauged}
matrix models, as can be seen in the following
example.
\end{itemize}

\begin{example} Take the simplest example of a gaussian measure $\rd\mu_N(x)=\rd\mu_N^{(0)}$, with a normalisation factor
given by 
\beq
\cN_N\triangleq \l[\mathrm{Vol}_g(\mathrm{U}(N))\r]^{-1}
\label{eq:NnvoluN}
\eeq
where the metric $g$ is the Ad-invariant metric induced by the Killing form on
$H_N$. The gaussian integration is trivial, and yields
\beq
\int_{H_N} \rd\mu_N^{(0)} = \l(\frac{2\pi t}{N}\r)^{N^2/2}
\eeq
The normalisation factor in \cref{eq:NnvoluN} is less trivially computed as a
product of volumes of spheres \cite{MR558859} and it takes the form of a Barnes double gamma
function:
\beq
\mathrm{Vol}_g(\mathrm{U}(N))=\frac{(2\pi)^{N(N+1)/2}}{G_2(N+1)}
\eeq
where $G_2(x+1)=\Gamma(x)G_2(x)$, $G_2(1)=1$. Computing the large $N$ asymptotics of
$\ln Z_N$ is an exercise in the use of the all-order Stirling formula for the
Gamma function; the result is
\beq
\ln Z_N(t)= \frac{N^2}{2}\l(\log(t)-\frac{3}{2}\r)-\frac{1}{12}\log
N+\zeta^{'}(-1)+\sum_{g\geq 2} \frac{B_{2g}}{2g(2g-2)} N^{2-2g}.
\label{eq:largeNgauss}
\eeq
We point out three features of this example.
\begin{enumerate}[i.]
\item In this case we can give a full answer to the question of finding a
  string dual for the matrix model. Indeed, \cref{eq:largeNgauss} coincides
  with the perturbative expansion of the $c=1$ string theory
  at the self-dual radius \cite{Gross:1990ub,Periwal:1993yu,Ghoshal:1995wm},
  upon identifying the string coupling constant $g_s$ and the cosmological constant $\mu$
  as $g_s=t/N$, $t=\ri\mu$.
\item The coefficients of $N^{2-2g}$ in the genus expansion of
  \cref{eq:largeNgauss} have a particular geometrical significance of their
  own: by the Harer--Zagier formula \cite{MR848681}, they coincide with the Euler characteristic of the
  moduli space of genus $g$-curves with no marked points and $g>1$,
\beq
\chi(\cM_{g,0})=\frac{B_{2g}}{2g(2g-2)}=\frac{(-1)^{g-1} 2^{2g-1}(2g-1)(2g-3)!}{2^{2g-1}-1}\int_{\overline{\cM}_{g,1}}\psi_1^{2g-2} \lambda_g
\eeq
This is perhaps the simplest setting where an answer to a topological problem
on the moduli space of Riemann surfaces (respectively, in the second equality,
an intersection theoretic problem on $\overline{\cM}_{g,n}$) is encoded into a
formal 1-matrix model\footnote{The distinction between {\it gauged} and {\it
    ungauged} matrix models is somewhat in the eye of the beholder: the same
  result would be obtained via the large $N$ study of the (ungauged) Penner matrix model
  in a double-scaling limit; see \cite{Distler:1990mt}.}. This is not an accident, and the rest of this
paper will be devoted to non-trivial extensions of this phenomenon.
\item A specular (in a precise sense) point of view to the previous statement is given by the study
  of the large~$N$ expansion in \cref{eq:largeNgauss} using {\it loop equations}
  \cite{Brezin:1977sv,Migdal:1984gj}; an infinite set of differential
  constraints on the cumulants of \cref{eq:w} (see \cite{Eynard:2015aea,Marino:2005sj} for reviews). Its upshot in our case is that the
  planar free energy, $\cF_0$, in \cref{eq:1/N} can be recovered as an integral
  transform of Wigner's semi-circle function,
\bea
\cF_0(t) &=& -t \int_{0}^{2\sqrt{t}} z^2 \rho(z)\rd z, \nn \\
\rho(z) &\triangleq & \frac{1}{2\pi\ri}\sqrt{z^2-4t}.
\eea
Equivalently, for $t\in\bbC^\star$ consider the {\it spectral curve}
$(C_t,\rd\lambda_t)$ given by the smooth, genus zero affine plane curve
$C_t=\{(z,y)\in \bbC^2 | y^2=z^2-4t\}$ endowed with the meromorphic
differential $\rd\lambda_t=ydz$. Then $F_0(t)$ is recovered from the rigid
special K\"ahler geometry relations \cite{Strominger:1990pd}
\beq
t=\frac{1}{2\pi \ri}\oint_A \rd\lambda, \qquad \frac{\de \cF_0}{\de t}=\frac{1}{2}\oint_B \rd\lambda, 
\eeq
where the integrals over $A\in H_1(C_t,\bbZ)$, $B \in H_1^{\rm BM}(C_t,\bbZ)$ denote respectively
a contour integral encircling the cut $[-2\sqrt{t},2\sqrt{t}]$ of $y(z)$ in
the $z$ plane, and the sum of principal value integrals from $-\infty^-$ to
$-2\sqrt{t}$ and from $-2\sqrt{t}$ to $-\infty^+$; here $-\infty^\pm$ refers
to the two pre-images of $z=-\infty$ under the branched covering map $z:C_t\to \bbC$.

\end{enumerate}

By the looks of it, Point ii) above relates the formal $1/N$ expansion of the gauged gaussian
matrix model to the computation of a particular class of intersection
numbers on the moduli space of curves; in physics language, these are
a subset of observables in a simple example of an {\it $A$-type} topologically twisted
string theory\footnote{In particular, these numbers are closely related to the
  equivariant Gromov--Witten theory of the affine line -- a deformation of
  topological gravity by linear insertions of Chern classes of the Hodge bundle.}. Point~iii) does the
same thing (at the leading order) for something superficially rather different:
special geometry governs the dependence on complex (vector) moduli of the
prepotential in type IIB compactification, and is precisely what is captured
by the planar limit of a {\it $B$-type} topologically twisted string theory. This
can actually be made more precise: it was shown in \cite{Ghoshal:1995wm} that
the $c=1$ string at the self-dual radius reproduces the genus expansion of the
topological A-model on the singular conifold, which is its own
self-mirror on the $B$-side.
\label{ex:gauss}
\end{example}

\subsection{A-model: intersection theory on $\overline{\cM}_{g,n}$}
The triangle of relations we found in this example between formal matrix models, intersection theory on
$\overline{\cM}_{g,n}$ and special geometry on a family of curves is part of a
more general story. For the general formally deformed measure $\rd \mu_N(x)$, we write
\beq
W_{g,h}(z_1,\dots,z_h; x) \triangleq \prod_{i=1}^hz_i^{-1}\sum_{k_1,\dots,k_h}\cW_{g,h;k_1,\dots,k_h}(x)\prod_{i=1}^h z_i^{-k_i}
\eeq
for the generating function of the cumulants of \cref{eq:1/N} at the $g^{\rm th}$ order in
the $1/N$ expansion of the formal 1-matrix model of \cref{eq:mun}. At the
leading (planar) order in $1/N$, the first loop equation reduces to
\beq
W_{0,1}(z) = \frac{1}{2t}P'(z) -\frac{1}{2t}M(z)\sqrt{\sigma(z)},
\label{eq:W01}
\eeq
where $\sigma(z)=(z-b_1(x,t))(z-b_2(x,t))$ and the {\it moment function}
$M(z;x)$ as well as the branchpoints $b_i(x,t)$ are
entirely determined by $P(z)$ and $t$. The second planar loop equation fixes
the two-point function to take the form \cite{Akemann:1996zr}
\beq
W_{0,2}(z_1,z_2) =
-\frac{1}{2(z_1-z_2)^2}+\frac{\sqrt{\sigma(z_1)}}{2(z_1-z_2)^2\sqrt{\sigma(z_1)}
\sqrt{\sigma(z_2)}}-\frac{\sigma'(z_1)}{4(z_1-z_2)\sqrt{\sigma(z_1)\sigma(z_2)}}.
\label{eq:W02}
\eeq
In two remarkable papers \cite{Eynard:2011kk,Eynard:2011ga} (see also \cite{DuninBarkowski:2012bw}), Eynard established the following
\begin{thm}
There exist explicit generating functions of tautological classes $\Lambda_{g,n}(x)\in
\bbC[[x]] \otimes R^\bullet(\overline{\cM}_{g,n})$,
$\cB_{g,n}(z_1,\dots,z_n;x) \in
\bbC[[x;z_1,\dots z_n]] \otimes R^\bullet(\overline{\cM}_{g,n})$ such that
\bea
W_{g,n}(z_1,\dots,z_n;x) &=&\int_{\overline{\cM}_{g,n}}\Lambda_{g,n}(x)
\cB_{g,n}(x,z_1,\dots,z_n), \nn \\
\cF_{g}(x)&=& \int_{\overline{\cM}_{g,0}}\Lambda_{g,0}(x).
\label{eq:fgmgn}
\eea
In \cref{eq:fgmgn}, the classes $\Lambda_{g,n}(x)$ and
$\cB_{g,n}(x,z_1,\dots,z_n)$ are entirely determined by knowledge of $W_{0,1}$
and $W_{0,2}$ in \cref{eq:W01,eq:W02}.
\label{thm:fgmgn}
\end{thm}

\cref{thm:fgmgn} encodes the solution of an intersection theoretic
problem on $\overline{\cM}_{g,n}$ into the knowledge of the correlators
$W_{g,n}(z_1,\dots,z_n;x)$ of an associated formal 1-matrix model. In fact,
the applicability of
\cref{thm:fgmgn} goes far beyond the realm of (multi-)matrix models: it
applies to any spectral curve setup 
and higher order correlators defined through the Eynard--Orantin topological
recursion as in the following section.

\subsection{B-model: the Eynard--Orantin topological recursion}

\begin{figure}[!h]
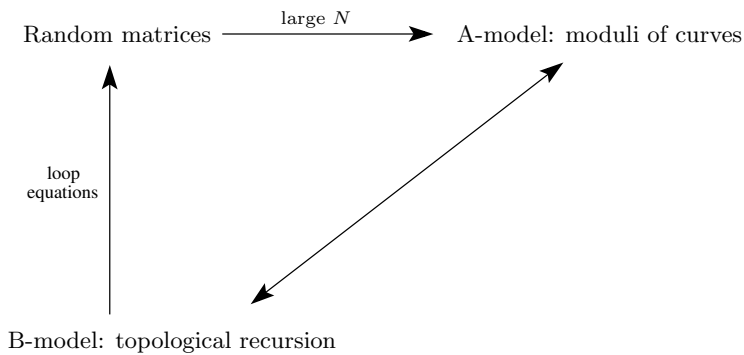
\caption{Large $N$ duality and mirror symmetry in $d=0$}
\label{fig:diagmm}
\end{figure}

\cref{thm:fgmgn} gives a means to compute the $1/N$ expansion of the correlators of the 1-matrix
model in terms of intersection numbers on $\overline{\cM}_{g,n}$, which in
general is hardly a simplification from the computational point of view. An independent way to
compute them was found in 2004 by Eynard, and further developed in subsequent
works with Chekhov and Orantin, in terms of a recursive solution of the loop
equations, as in the following
\begin{thm}[\cite{Eynard:2004mh,Chekhov:2005rr,Eynard:2007kz}]
Let $\overline{C}_t$ be the normalisation of the projective closure of the plane curve given by the zero locus of
$y^2-\sigma(z)$ in $\bbC^2$ with coordinates $(z,y)$, and write $\rd E_{w}(z) \in \Omega^1(\overline{C}_t\setminus \{w=z\})$ 
for the logarithmic derivative of the prime form on $\overline{C}_t$,
normalised as $\oint_A \rd E_{w}=0$; here $A$ is an oriented loop winding once
counter-clockwise around the segment $[b_1(x,t), b_2(x,t)]$. Then, for $2g-2+h>0$ the following recursions hold:
\begin{align}
W_{g,h+1}(z_0, z_1 \ldots, z_h) =& \sum_{i=1,2}  \underset{z=b_i}{\rm Res~}
\frac{\rd E_{z_0}(z)}{2 M(z) \sqrt{\sigma(z)}} \Big ( W_{g-1,h+2} (z, \overline{z}, z_1, \ldots, z_{h} )\nn \\
+& {\sum_{l=0}^g}  {\sum'_{J\subset H}} W^{(g-l)}_{|J|+1}(z, z_J)
W^{(l)}_{|H|-|J| +1} (\overline{z}, z_{H\backslash J}) \Big),
\label{eq:toprecW}
\end{align}
\beq
\cF_g = \sum_{i=1,2}  \underset{z=b_i}{\rm Res~}
\frac{\rd E_{z_0}(z)}{2 M(z) \sqrt{\sigma(z)}} \int^z \rd z' W_{g,1}(z')
\label{eq:toprecF}
\eeq
Here $I \cup J=\{z_1,\dots, z_h\}$, $I \cap J=\emptyset$,
$\sum'$ denotes omission of the terms $(h,I)=(0,\emptyset)$ and $(g,J)$, and
the primitive in \cref{eq:toprecF} is independent of its base point.
\end{thm}
\cref{eq:toprecF,eq:toprecW} together make up the {\it topological recursion}: they determine recursively
all the higher orders of the correlators $W_{g,h}$ starting from
$(g,h)=(0,1), (0,2)$ in \cref{eq:W01,eq:W02}, and consequently the all-order
expansion of the partition function. Its frame of applicability goes far
beyond the formal analysis of single-cut 1-matrix models, encompassing also
multi-cut solutions, multi-matrix models, and non-polynomial potentials. The
overall picture that emerges can be codified by the diagram of \cref{fig:diagmm}.

\label{sec:0d}

\section{The GOV correspondence for $S^3$}

How much of this story is specific to matrix models? It should be noticed that
't Hooft's original argument in no way depended on the dimensionality of
the gauge theory we are doing perturbation theory on, as long as the
 interacting theory has a weakly coupled limit given by a free Gaussian theory, a
  sensible UV-regularisation scheme exists, and renormalisation does not lead to
  contributions to the perturbation series which are individually factorially
  divergent at each order in the topological
  expansion. A natural, if perhaps bold, question that could be asked is then: ``Can the top left
corner in \cref{fig:diagmm} be replaced by a higher dimensional
(topological) gauge theory?''. 

Evidence that this is not a mere speculation
comes from a non-trivial example, found by Gopakumar, Ooguri and Vafa in
\cite{Gopakumar:1998ki, Ooguri:1999bv}, where the {\it entire} setup
of \cref{fig:diagmm} carries forth verbatim to higher dimension. Let $M$ be a smooth, closed,
oriented real 3-manifold and $\cK$ be a link in it. For a smooth $\mathrm{U}(N)$ gauge connection
$\cA$ on $M$ let
\beq
\mathrm{CS}[\cA]\triangleq \int_M \mathrm{Tr}_\square \l( \cA
\wedge \rd \cA + \frac{2}{3} \cA \wedge \cA \wedge \cA \r) 
\eeq
be the Chern--Simons functional of $\cA$. It was Witten's realisation that the formal functional integrals
\cite{Witten:1988hf} 
\bea
Z_N^{\rm CS}(M, k) & \triangleq &\int D[\cA] \exp\l[  \frac{\ri k}{4\pi} \mathrm{CS}[\cA]  \r],
\nn \\
W^{\rm CS}_{N,\rho}(M, \cK, k) & \triangleq & Z_N^{-1}\int  D[\cA]
\tr_\rho\l(\mathrm{Hol}_\cK(\cA)\r) \exp\l[  \frac{\ri k}{4\pi} \mathrm{CS}[\cA]\r] 
\label{eq:CSvev}
\eea
lead to (framed) topological invariants of $M$ and $\cK$ in the form of the
Reshethikin and Turaev's $\mathrm{sl}_N$-quantum invariants; here $k \in \bbZ^*$ is
the Chern--Simons level, and the coloring $\rho$ is an irreducible representation of
$\mathrm{U}(N)$. Their large $N$ expansion can be worked out entirely explicitly in different
ways, for example by resorting to the surgery techiniques of
\cite{Witten:1988hf}: for the partition function we find
\beq
\ln Z_N^{\rm CS}(M, k) = \sum_{g \geq 0} g_s^{2g-2} \cF^{\rm CS}_{g}(t)
\eeq
where $g_s=2\pi \ri (k+N)^{-1}$, $t= g_s N$, and
\beq
\cF_g^{\rm CS}(t) = \frac{|B_{2g}|}{2g(2g-2)!} \mathrm{Li}_{3-2g}(e^{-t})+\frac{(-1)^g
  B_{2g} B_{2g-2}}{2g (2g-2)(2g-2)!}
\label{eq:FgCSS3}
\eeq
for $g\geq 2$, with similar formulas for $g=0,1$ \cite{Gopakumar:1998ki}. If
we trust that a higher dimensional analogue of the discussion of the previous
section exists, the two questions we are tasked to answer are:
\ben
\item characterise the large $N$ A-model dual of Chern--Simons theory in terms
  of a precise intersection theoretic problem on $\overline{\cM}_{g,n}$;
\item characterise its mirror, large $N$ B-model dual in terms of the
  topological recursion on a specific spectral curve setup.
\een

\subsection{A-model}

The A-model dual found in \cite{Gopakumar:1998ki} is a sort of $q$-deformation
of the setup of \cref{ex:gauss}, and it has an explicit presentation in terms
of an intersection theory problem on a moduli space of stable {\it maps}. First
off, it was found in \cite{Witten:1992fb} that the partition function of $\mathrm{U}(N)$ Chern--Simons theory on any closed three
manifold $M$ is equal to the partition function of the {\it open} topological A-model
on the total space of the cotangent bundle $T^*M$, with $N$ Lagrangian
A-branes wrapping its zero section. The proposal of
\cite{Gopakumar:1998ki} is to relate the latter at large $N$ to the {\it ordinary, closed} A-model/Gromov--Witten theory on a
target space obtained from $T^*S^3$ via a complex deformation to a normal
singular variety
and a bimeromorphic resolution of its singularities (the {\it conifold
  transition}), as follows. When $S^3$ has radius one in the canonical metric,
we have an obvious isometry $T^* S^3 \simeq
\bbR^3 \times S^3 \simeq \mathrm{SL}(2, \bbC)$ given by the decomposition of
a special linear matrix $A$ into radial (positive definite) and polar (unitary)
part: 
\beq
A = U \re^{H}, \qquad U\in \mathrm{SU}(2), \quad H \in
\mathcal{H}_0(2,\mathbb{C}).
\eeq
In particular
this endows $T^* S^3$ with a complex structure given by its presentation as a
quadric hypersurface $\det A=1$ in $\mathrm{Mat}(2,\bbC) \simeq \bbC^4$. 

We now perform the following two operations on this A-model target space:
\ben
\item First we vary the
radius of the base unit sphere to give a flat family $\psi: X = {\rm GL}(2,\mathbb{C}) \to \bbC^*$ via
the determinant map, whose fiber $X_{[\mu]}$ at a point $\mu$ with
${\rm Im}\,\mu =0$ and ${\rm Re}\,\mu>0$ is isomorphic to the cotangent bundle
$T^*S^3_{[\mu]}$. Notice that we are doing {\it nothing here} either from
the point of view of Chern--Simons theory or the open topological A-model on
the cotangent bundle, as a homeomorphism of the base (resp. a complex
deformation of the total space) leave invariant the CS partition function
(resp. the A-model partition function). 
\item The second is to add the locus of non-invertible matrices to form:
\beq
\tilde \psi\,:\, {\rm Mat}(2,\mathbb{C}) \longrightarrow \bbC\,.
\eeq
The singular fiber $X_{[0]}$ 
above $\mu = 0$ is the singular quadric $\det A=0$: this is the singular
conifold of \cref{ex:gauss}. The latter admits a canonical
toric minimal resolution 
\beq
\pi\,:\,\widehat{X} \longrightarrow X_{[0]},\qquad  \widehat{X} \triangleq \big\{(\rho(A),v) \in X_{[0]} \times \mathbb{P}^1,\quad \rho(A) v =0\big\},\qquad 
\label{eq:rescon} 
\eeq
where $\pi$ is the projection to the first factor. The point $A = 0$ is
singular in $X_{[0]}$, and its fiber is a complex projective line with
$[v_1:v_2]$ as homogeneous coordinates. Using coordinate charts on
$\mathbb{P}^1$ exhibits $\widehat{X}$ as the total space of $\cO(-1)\oplus
\cO(-1)\to \mathbb{P}^1$, i.e. the {\it resolved conifold}.
\een
The proposal of \cite{Gopakumar:1998ki} is that
\beq
\cF_g^{\rm CS}(t) = {\rm GW}_g^{\widehat X}(t)
\label{eq:GOVS3}
\eeq
where we have identified the 't Hooft parameter on Chern--Simons theory with
the A-model K\"ahler parameter dual to the curve class of the base $\cO(-1)^{\oplus
  2}_{\bbP^1}$. In \cref{eq:GOVS3}, ${\rm GW}_g^{\widehat X}$ denotes the
genus-$g$ primary Gromov--Witten potential of $\widehat{X}$, 
\beq
{\rm GW}_g^{\widehat X}(t) = \sum_{d \geq 0}\re^{t
  d}\int_{[\overline{\cM}_{g,0}(\widehat X,d)]^{\rm vir}} 1
\eeq
where $\overline{\cM}_{g}(\widehat X,d)$ is the stack of degree $d$ stable
maps from genus $g$ curves to
$\widehat X$ and $[\overline{\cM}_{g}(\widehat X,d)]^{\rm vir}$ denotes its virtual fundamental cycle.

There are a number of
reasons, at various degrees of rigour, to take this proposal seriously:
\ben
\item Firstly, it holds true asymptotically in the $t\to 0$ limit:
  indeed on the CS
  side, the partition function is known to reduce in this limit to the gauged gaussian
  matrix model of \cref{ex:gauss} \cite{Periwal:1993yu}; and on the GW side,
  the $t\to 0$ regime
  corresponds to the small volume limit (the {\it singular conifold}). The
  claim then follows from the conifold/$c=1$ duality we discussed in \cref{ex:gauss};
\item There are also heuristic physics arguments in favour of the duality. One
  hinges on viewing \cref{eq:GOVS3} as an open/closed duality in the
  A-model, the equality can be interpreted as a brane/flux duality in the
  physical type IIA superstring: in principle, both sides compute superpotential observables in
  effective $\cN=1$ theories, obtained either as the world-volume theory of wrapped
  $D6$-branes in $T^*S^3$ of via
  turning on internal RR field strengths on the resolved side \cite{Bershadsky:1993cx,Vafa:2000wi}.
These string configurations can then be related by lifting
  them to M-theory \cite{Atiyah:2000zz}.
\item A further microscopic derivation of the duality can also be given \cite{Ooguri:2002gx} upon relating
  the open partition function on the deformed conifold and the closed
  partition function on the resolved conifold by their UV completion in terms
  of a gauged linear $\sigma$-model: here the $t\to 0$ limit leads 
  to a coexistence of Coulomb/Higgs phases on the closed string world-sheet, with
  the holes of the open theory arising from integrating out the contribution
  of Coulomb regions.
\item What is more, the equality of \cref{eq:GOVS3} is a sharp mathematical
  statement about RTW invariants and GW invariants that can be rigorously proven, or
  disproven, by an explicit calculation. The l.h.s. can be rigorously shown to
  lead to the expansion of \cref{eq:FgCSS3} from an explicit MacLaurin
  expansion of the finite sums appearing in the surgery formula for the
  Reshetikhin--Turaev invariant \cite{Gopakumar:1998ki}; and on the other hand
  the r.h.s. is
  amenable to a direct localisation analysis in Gromov--Witten theory, first
  performed in \cite{MR1666787,MR1728879}. The end result is an exact agreement
  of the two generating functions.
\item Finally, a natural question is how the story generalises when we
  incorporate links in Chern--Simons theory. It was first proposed in
  \cite{Ooguri:1999bv} that when $\cK=\bigcirc$ is the trivial knot,
  \cref{eq:GOVS3} should generalise to an equality between the colored HOMFLY
  invariants of \cref{eq:CSvev} and {\it open} GW potentials for a suitable
  choice of Lagrangian A-branes $L$:
\beq
W_{g,h, \vec{d}}^{\rm CS}\l(S^3, t, \bigcirc\r) = \mathrm{GW}^{\widehat X,L}_{g,h,\vec{d}}(t)
\eeq
where in the l.h.s. we consider a knot invariant obtained as a power
sum (instead of Schur functions, as in \cref{eq:CSvev}) holonomy invariant specified by a vector
of integers $d_1, \dots, d_h$, we take its connected part, and we expand it at large
$N$ as in \cref{eq:1/Nw}, and in the r.h.s. we take the generating function of open GW invariants
with boundary on the fixed locus of a real involution (a real bundle on the
equator of the base $\bbP^1$), with fixed genus $g$ and number of holes $h$
for the source of curves and winding numbers $d_1, \dots, d_h$ around $S^1
\subset \bbP^1$. Indeed, the line of reasoning of \cite{Vafa:2000wi} carries through to this
setting, and also in this case a localisation definition/computation of open
GW invariants can be performed \cite{Katz:2001vm}, confirming the prediction
of \cite{Ooguri:1999bv}.
\een

\subsection{B-model}

\begin{figure}[t]
\begin{minipage}{0.49\linewidth}
\centering
\vspace{0pt} 
\includegraphics[scale=0.75]{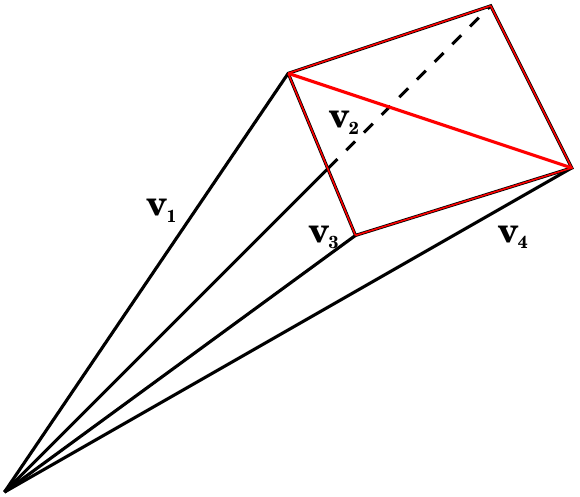}
\vspace{0pt} 
\label{fanconifold}
\end{minipage} 
\begin{minipage}{0.49\linewidth}
\centering
\vspace{0pt} 
\includegraphics{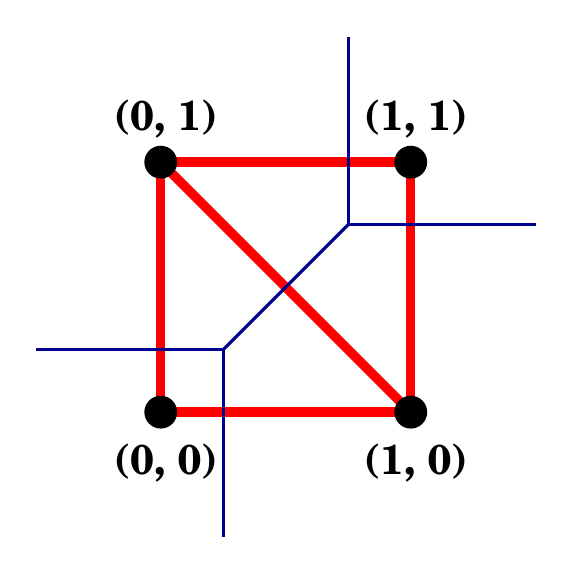}
\vspace{0pt} 
\label{tordiagconifold}
\end{minipage} 
\caption{The fan (left) of the resolved conifold: its skeleton is given by
  four rays, labelled $v_i$ in the picture, whose tips lie on an affine hyperplane at unit distance from the
  origin of $\bbZ^3$. The intersection of the fan with the hyperplane is shown
on the right (the {\it toric diagram}); superimposed is the pq-web diagram.}
\label{fig:fan}
\end{figure}

Since the A-model target space is a (non-compact) Calabi--Yau threefold, we
would expect a geometric mirror picture in terms of rigid special K\"ahler
geometry of a family of local CY3s, akin to what happened in \cref{ex:gauss}
for the case of matrix models. Not only is this case, but the analogy with the
spectral curve setup of random matrices is even more poignant -- the special
geometry relations on the local CY3 mirror (which, in this particular case, is just the
deformed conifold geometry of the previous section) can be shown to reduce to
Seiberg--Witten-type relations on a family of genus zero spectral curves, given by the
Hori--Iqbal--Vafa mirror of $\widehat{X}$:
\beq
t=\frac{1}{2 \pi \ri}\oint_A \ln y d\ln x, \qquad \frac{\de
  \mathrm{GW}^{\widehat X}_0}{\de t} = \frac{1}{2} \oint_B \ln y d\ln x
\eeq
where $A,B$ are homology 1-cycles relative to the principal divisors $x=0$,
$y=0$ of the plane curve defined by $P_{\widehat X}(x,y)=1+x+y+\re^{t}=0$; here $P_{\widehat
  X}$ is the Newton polynomial of the diagram in
\cref{fig:fan}. What is more, motivated by the study of the chiral boson theory of
\cite{Aganagic:2003qj}, it was first suggested in \cite{Marino:2006hs}, and then postulated in full detail
in \cite{Bouchard:2007ys}, that the higher genus B-model potentials (and, by mirror symmetry,
the higher genus open GW invariants of $\widehat{X}$) should be computed by the
topological recursion of \cref{eq:toprecW,eq:toprecF} upon identiying
$ydx=W_{0,1}(x)$. This prediction has been subsequently proved in full rigour
in \cite{Eynard:2012nj} for all toric Calabi--Yau manifolds that are
symplectic quotients\footnote{See \cite{Fang:2016svw} for a recent generalisation of
  this result to the case of orbifolds.}. 

\label{sec:S3}

\section{The GOV correspondence for Clifford--Klein 3-manifolds}
We have seen how the picture of \cref{fig:diagmm} generalises verbatim to the
higher dimensional 
setting of Chern--Simons theory on $S^3$: the top-right corner has a
concrete A-model picture in terms of the Gromov--Witten theory of the resolved
conifold, and the bottom-left corner is encoded  by the topological recursion on
its Hori--Iqbal--Vafa mirror curve. The emerging picture is not only beautiful
and unexpected; from a practical point of view, it has deeply affected the relation between quantum invariants, curve-counting
invariants, and the topological recursion. 
Appealing and impactful as it has been, this is however just {\it one} example
-- placed
in the classical setting of $\mathrm{U}(N)$ theories, encompassing the simplest closed
3-manifold, and with the simplest knot therein. It is then
natural to seek a strict generalisation of the GOV correspondence to
other\footnote{The following list leaves out one important generalisation,
  tying together categorification on the knot theory side \cite{Gukov:2004hz}, refined BPS
  counting on the A-model side \cite{Iqbal:2007ii,Katz:1999xq,Choi:2012jz},
  and some version of the $\beta$-deformation on the
  B-model side \cite{Brini:2010fc,Aganagic:2011mi,Aganagic:2011sg}.}
\ben
\item {\bf knots}: in its strictest sense, the GOV correspondence has been
  shown to carry
  through to the case of torus knots both in terms of an enumerative theory of
  open GW invariants on the A-side \cite{Diaconescu:2011xr} and the
  topological recursion on a specific spectral curve setup on the B-side
  \cite{Brini:2011wi}. For general knots, it would seem that substantially new
  ideas are needed both on the A- and the B-side of topological string theory
  \cite{Aganagic:2013jpa};  the role of the topological recursion in
  particular is however less clear in this setting \cite{Gu:2014yba}.
\item {\bf (classical) gauge groups}: SO($N$) and Sp($N$) Chern--Simons theory
  at large $N$, which compute in particular the Kauffmann invariant of links, can also fit in the picture of the previous section by
  considering suitable orientifolds of the resolved conifold \cite{Sinha:2000ap}, for which an
  operative definition of unoriented invariants can be given either by
  localisation or via the topological vertex \cite{Bouchard:2004iu,Bouchard:2004ri}.
\item {\bf 3-manifolds}: this is possibly the boldest generalisation --
  replace altogether $S^3$ by an arbitrary closed 3-manifold $M$. At face
  value this boils down to solving the problem for arbitrary knots on $S^3$:
  the partition function of a link $\LL$ in $M$ would then be recovered by the
  surgery formula from the partition function of a 2-component link $\LL \sqcup \LL_M$ in $S^3$,
  where $\LL_M$ is a framed link $M$ can be obtained from (this always exists by
  Lickorish's theorem). However, both the surgery formula in Reshetikhin--Turaev theory
  \cite{MR1875611,Witten:1988hf}, and functional localisation in Chern--Simons
  theory \cite{Beasley:2005vf, Kallen:2011ny,Blau:2013oha} lead to an
  expression of the partition function as a sum over contribution labelled by
  flat connections (classical vacua) on $M$:
\beq
Z^{\rm CS}_N(M,k) = \sum_{\mathsf{v}\in \mathrm{Hom}(\pi_1(M), \mathrm{U}(N))/\mathrm{U}(N)} Z^{\rm CS}_{N,\mathsf{v}}(M,k).
\label{eq:sumflat}
\eeq
\een
It was first proposed in \cite{Aganagic:2002wv} that the finer invariants
given by the {\it individual summands} $Z^{\rm CS}_{N,\mathsf{v}}$ may also be interpreted as the A/B-model partition
function on a background specified by $\mathsf{v}$; notice that this is a more
refined object to deal with than the partition function of the link $\LL_M$,
where all these contributions are summed over. That a dual curve counting
theory exists is encouraged by the
successful test of this proposal for the case of $L(p,1)$ lens spaces in
\cite{Aganagic:2002wv,Halmagyi:2003ze}. The case of more general 3-manifolds was
considered in \cite{iolpq,Borot:2014kda,Borot:2015fxa,Brini:2017gfi}, and we review it below.

\subsection{CS theory on Clifford--Klein 3-manifolds}
We start by recalling the following
\begin{defn}
A {\rm Clifford--Klein} 3-manifold $(M,g)$ is a closed oriented smooth 3-manifold
$M$ admitting a smooth metric $g$ of everywhere strictly positive Ricci curvature.
\end{defn}
Equivalently, by Hamilton's theorem, it is a spherical space form,
$M=S^3/\Gamma$ for $\Gamma$ a freely acting finite isometry group of $S^3$
w.r.t. its canonical metric; and by Perelman's elliptisation theorem, it
follows that these coincide with the orientable 3-manifolds having finite
fundamental group. The classification of the possible
$\Gamma$ goes back to Hopf \cite{MR1512281}: these are central extensions of
the left-acting finite subgroups of ${\rm SL}_2(\bbC)$, which admit an ADE
classification; see \cite[Appendix~A]{Borot:2015fxa} for more details. We
restrict henceforth for simplicity of exposition to the case where the central
extension is trivial and $\Gamma$ is one such ${\rm SL}_2(\bbC)$-subgroup; most of
our arguments to follow will be unaffected by this.

Now since $|\Gamma|<\infty$, the sum in \cref{eq:sumflat} truncates at finite
$N$. Denote by $\mathfrak{V}_{\Gamma,N}$ the finite set of gauge-equivalent
flat connections and by $\mathfrak{V}_{\Gamma} = \lim_{N \rightarrow \infty}
\mathfrak{V}_{\Gamma,N}$ be its direct limit with respect to the
composition of morphisms given by the embedding ${\rm U}(N)
\hookrightarrow {\rm U}(N + 1)$. Then 
\beq
\mathfrak{V}_{\Gamma} = \mathbb{N}^{R + 1},\qquad \mathfrak{V}_{\Gamma,N} = \l\{(N_0,\ldots,N_R) \in \mathbb{N}^{R + 1},\quad N_0 + \sum_{i = 1}^R D_iN_i = N\r\}.
\eeq
where $R$ is the number of nodes in the simply-laced Dynkin diagram associated
to $\Gamma$, and $D_i$ are the respective Dynkin indices. When $N \rightarrow
\infty$, we thus consider a CS vacuum
$[\mathcal{A}]_{t}$ parametrized by $t \triangleq
N_i g_s$ for $i \in \{0,\dots, R\} $, and in particular the rank is
encoded in the 't Hooft parameter $t = N g_s$. The resulting partition
functions at $N=\infty$ have a standard $1/N$ expansion
\beq
\ln Z^{\rm CS}_{N,\mathsf{v}(t)}(g_s)=\sum_{g\geq 0}g_s^{2g-2}\cF_g^{\rm CS}(t_0,
\dots, t_R)
\label{eq:ZCSGamma}
\eeq
with free energies depending now on $R+1$ 't~Hooft parameters. 

These manifolds
also carry a natural class of knots with them. For $S^3$, the standard GOV
correspondence focused on the unknot -- which could be regarded as the fibre knot
of the Hopf fibration on $S^2$. Now Clifford--Klein manifolds are also
Seifert-fibred, and in the ADE case they can be regarded as Seifert fibrations
over an ADE $\bbP^1$-orbifold with one (resp. three) exceptional fibres for
Dynkin type A (resp. D and E). As for $S^3$, we will similarly be interested in the $1/N$ expansion of
the RTW invariant for the knots $\cK_f$ running around these exceptional fibres:
\beq
W^{\rm CS}_{N,\mathsf{v}(t)}(\cK_f,g_s,\vec d)=\sum_{g\geq
  0}g_s^{2g-2}\cF_{g,h}(t_0, \dots, t_R; \vec d)
\label{eq:WCSGamma}
\eeq

The quest is to find now a Calabi--Yau threefold geometry $\widehat{X}_\Gamma$
with special Lagrangians $L_\Gamma$, as well as a
spectral curve setup $\Sigma_\Gamma$ for each such $\Gamma$, such that
open/closed GW theory on $(\widehat{X}_\Gamma,L_\Gamma)$ and the topological
recursion on $\Sigma_\Gamma$ lead to \cref{eq:ZCSGamma,eq:WCSGamma}. This
program was completed in \cite{Borot:2015fxa,Brini:2017gfi}, generalising ideas on the lens space case in \cite{Aganagic:2002wv}.

\subsection{A-model from geometric transition}

On the A-model the idea is fairly simple: we take seriously the geometric
transition argument of \cref{sec:S3} and apply it to the setting at hand. To
this aim, notice that the $\Gamma$-action on the resolved conifold $\widehat{X}$ only acts on the first factor of
\cref{eq:rescon}, giving a fibrewise action on $p\,:\,\widehat{X}
\rightarrow \mathbb{P}^1$ (the second factor in \cref{eq:rescon}); the fibre
over a point $z \in \mathbb{P}^1$ is isomorphic to a Gorenstein surface
singularity of the same ADE type of $\Gamma$. The topological A-string on this
target geometry can be studied in several phases, two of which are
distinguished: 
\bit
\item in the {\it orbifold} chamber, we are looking at a theory of twisted
  stable maps on the Calabi--Yau stack $\widehat{X}^\Gamma_{\rm orb}\triangleq
  [\widehat{X}/\Gamma]$. This is the maximally singular phase containing the
  $\Gamma$-orbifold of the conifold point, which is the natural point of
  expansion for the dual Chern--Simons theory;
\item in the {\it large radius} chamber, we take a crepant resolution
  $\widehat{X}^\Gamma_{\rm res}$ of the singularities of (the coarse moduli space of)
$\widehat{X}^\Gamma_{\rm orb}$ obtained by canonically resolving the surface singularity
$\bbC^2/\Gamma$ fibrewise, and we are looking at the ordinary GW theory of
  $Y^\Gamma_{\rm res}$.
\eit

We have four remarks about the resulting target space geometry. 

\begin{description}
\item[R1] in either case the target supports at least a $(\bbC^\star)^2$ torus
  action; this is the product of the 1-torus action that rotates the base
  $\bbP^1$ and the fibrewise 1-torus action inherited from the scalar action
  on the $\bbC^2$-fibre of $\widehat{X}$. In particular, as for the usual
  toric case, we will specialise to a resonant 1-subtorus
 by imposing that its action is trivial on the canonical bundle
  of the target; this acts with compact fixed 0- and
  1- dimensional fixed loci.  This in turn allows to define equivariant GW invariants by localisation on either
  $\widehat{X}^\Gamma_{\rm res}$ or  $\widehat{X}^\Gamma_{\rm orb}$.
\item[R2] As we already mentioned, the $\Gamma$-action is fibrewise and it covers the
  trivial action on the base $\bbP^1$; this circumstance will be important in
  a moment.
\item[R3] For all $\Gamma$, there are natural anti-holomorphic involutions
  whose fixed loci define Lagrangians for both
  $\widehat{X}^\Gamma_{\rm res}$ and $\widehat{X}^\Gamma_{\rm orb}$. These
  generalise the toric Lagrangian branes of
  \cite{Aganagic:2000gs,Ooguri:1999bv,MR2861610}, and the residual Calabi--Yau
  $\bbC^\star$ action of {\bf R1} above allows to define/compute open GW
  invariants by localisation. In particular, the methods of \cite{MR2861610}
  carry through verbatim to the setting at hand.
\item[R4] The anti-holomorphic involution of \cite{Sinha:2000ap} turns out to
  commute with the $\Gamma$-action, for all ADE types, and an orientifold
  theory can be defined in the same way. 
\end{description}

\subsection{B-model from geometric engineering}
{\bf R1-R2} above define completely what would be the top-right (A-model)
corner of a diagram like \cref{fig:diagmm} for the case at hand.  Now notice
that in {\bf R1} above, for type D and E the torus action we highlighted
does not extend to a full three-torus action on $\widehat{X}^\Gamma$ since
$\Gamma$ is non-abelian. In particular, we cannot resort to the toric mirror
symmetry methods of \cite{Hori:2000kt,Hori:2000ck} to find a spectral curve
setup for these cases, unlike for type A. A way out on physical grounds is
however pointed at by {\bf
  R2}: since the action is fibrewise, the topological A-string on these
backgrounds is known to geometrically engineer in a suitable limit a
4-dimensional $\cN=2$ gauge theory with simply-laced compact gauge group corresponding
to the Dynkin type of $\Gamma$ (and no adjoint hypermultiplets, since $\bbP^1$
has genus zero) \cite{Katz:1999xq}; so it is expected 
in this degenerate 
limit, which corresponds to a divisor at infinity in the stringy K\"ahler moduli space of
$\widehat{X}^\Gamma$, to have a mirror picture in terms of spectral curves
of Seiberg--Witten type \cite{Katz:1999xq,Seiberg:1994rs}. As a matter of
fact, even away from
this limit the A-model is still expected to give rise to a gauge theory with
eight supercharges, albeit in one dimension higher -- namely $\cN=1$ pure super
Yang--Mills theory on $\bbR^4\times S^1$, with the ``field-theory limit'' of
\cite{Katz:1999xq} corresponding to the fifth-dimensional circle shrinking to
zero-size. 

\begin{figure}
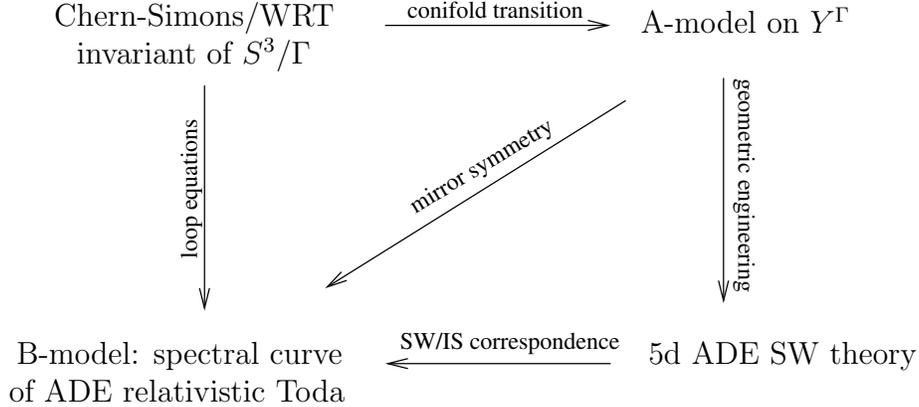
\caption{The GOV correspondence for $S^3/\Gamma$.}
\label{fig:diaggamma}
\end{figure}

Now for type $\mathrm{A}_N$, the five-dimensional theory also enjoys a description in terms of spectral
curves of Seiberg--Witten/Hori--Iqbal--Vafa type: these are the spectral
curves of the periodic Ruijsenaars system (relativistic Toda chain) with $N$
sites \cite{Nekrasov:1996cz}; the 4d limit corresponds to the non-relativistic limit. Furthermore, it has long been known that for all ADE types the relevant
SW curves should coincide with the spectral curves of the Lie-theoretic generalisation
of the non-relativistic Toda chain \cite{Martinec:1995by} specialised to ADE
Lie algebras. Putting all this together, it is then natural to look for a relativistic deformation of
\cite{Martinec:1995by} to supply the spectral curves defining a candidate
B-model mirror for the non-toric geometries of the previous section, and these
can be computed from the setup of \cite{Fock:2014ifa,Williams:2012fz}. We
refer the reader to \cite[Section~2]{Brini:2017gfi} for a more detailed
review; the upshot is that the spectral curves can be computed in the
following two steps:
\ben
\item fix $\rho$ to be a minimal irreducible $\cG_\Gamma$-module, where
  $\cG_\Gamma$ is the simply-connected simple Lie group over $\bbC$ of ADE
  type specified by $\Gamma$, and consider the characteristic polynomial of a
  group element $g$ in the representation $\rho$
\beq
\mathfrak{p}^\Gamma_\rho(g) \triangleq \det_\rho(\mathrm{id}~x - g) : \cT_\Gamma/{{\rm
    Weyl}(\cG_\Gamma)} \to \bbC[x]
\eeq
Here $\cT_\Gamma$ is the Cartan torus of $\cG_\Gamma$. We can decompose this on the Weyl character ring upon expanding the
determinant in a polynomial with coefficients given by anti-symmetric
characters of the representation $\rho$, and then write the latter as
polynomials of the fundamental characters $u_i$:
\beq
\mathfrak{p}^\Gamma_\rho(g) = \sum_{n=0}^{\dim \rho}(-1)^{\dim \rho-n}
\chi_{\wedge^{\dim \rho-n} \rho}(g) x^n, \qquad  \chi_{\wedge^n} \in
\bbZ[u_1, \dots, u_R]
\eeq
\item Now let $u_{\mathsf{k}}$ be the character of the maximally dimensional
  fundamental representation\footnote{In two cases there is an ambiguity which
    is resolved as follows. For type $\mathrm{E}_6$, pick either of
    the $\mathbf{27}$ or $\overline{\mathbf{27}}$. For type $\mathrm{A}_{2n}$,
    \cref{eq:Bcurve} is modified by shifting $u_i\to u_i+\delta_{i,n}y+ \delta_{i,n+1}u_0y^{-1}$.}. Then the spectral curve is
defined by the family of non-compact Riemann surfaces given by the polynomial equation
\beq
\mathfrak{p}^\Gamma_\rho\l(x; u_i+\delta_{i,\mathsf{k}}(y+u_0y^{-1}\r)=0
\label{eq:Bcurve}
\eeq
along with a choice of recursion differential given by the Poincar\'e one form
$W_{0,1}=\ln y \rd \ln x$, and higher order generating functions obtained by
the topological recursion on \cref{eq:Bcurve}. 
\een
The resulting web of relations is pictured in \cref{fig:diaggamma}.

\subsection{The LMO invariant and the topological recursion}
Our proposal fills all the vertices in the diagram of \cref{fig:diaggamma}:
the $\mathrm{sl}_N$-Reshetikhin--Turaev knot invariants of fibre knots in $S^3/\Gamma$ in the
$1/N$ expansion around a fixed flat connection would equate the open Gromov--Witten potentials of
$(\widehat{X}^\Gamma, L^{\Gamma})$ on one hand, and the Eynard--Orantin
invariants of \cref{eq:Bcurve} on the other; a similar statement applies to
the RT invariant of $S^3/\Gamma$ itself, the closed GW potential of
$\widehat{X}^\Gamma$, and the topological recursion free energies. 

However this is all conjectural for the moment -- so how do we prove this? One way to proceed is to resort to the relation of
Chern--Simons theory on Seifert fibred spaces (such as the Clifford--Klein
manifolds) and matrix models. It was shown by Mari\~no in
\cite{Marino:2002fk} that Witten's surgery formula for these spaces allows to
rewrite the CS partition function as well as the Wilson loops around certain
classes of knots as a matrix integral; in particular, the quantum invariants
\cref{eq:ZCSGamma,eq:WCSGamma} of $S^3/\Gamma$ and fibre knots therein around
the exceptional fibres fall squarely in this category
\cite{Marino:2002fk,Beasley:2005vf,Blau:2013oha}. When $\mathsf{v}=0$ is the
trivial flat connection, the resulting matrix model turns out to be a trigonometric
deformation of the gauged gaussian matrix model of \cref{ex:gauss}: it is a
canonical ensemble with gaussian 1-body interaction and a sum of $q$-deformed
Vandermonde 2-body interactions, with coefficients determined by the orders of the
exceptional fibres of the Seifert fibration
\cite{Marino:2002fk,MR2054838,Brini:2011wi,Borot:2014kda}: this restriction gives rise
to the so-called LMO invariant. 

This presentation is amenable to a
large $N$ analysis via loop equations, akin to that of Point iii) in
\cref{ex:gauss}, leading for all $\Gamma$ to a singular integral equation to
be solved by the input datum of the recursion -- the planar disk function
$W_{0,1}$. Such an analysis was performed in
\cite{Brini:2011wi,Borot:2014kda,Borot:2015fxa,Brini:2017gfi}; the strategy, and ensuing
results are as follows.
\ben
\item It can readily be shown that these are single-cut matrix models due to
  the gaussian nature of the 1-body potential, and that the (exponentiated)
  planar resolvent $W_{0,1}(x)$ is
  never a log-algebraic function of its argument except when $\Gamma$ is the
  trivial group. However, a strategy introduced in \cite{Brini:2011wi} for
  torus knots and then employed on a full scale on \cite{Borot:2014kda} is to
  considered a symmetrised version of the resolvent, which leads to the same
  large $N$ eigenvalue density as the original resolvent on the physical cut
  and may be such that the sheet transitions given by crossing the cuts close to
  a finite group. 
\item The previous step can be performed for any Seifert 3-manifold as that is
  the level of generality to which the ideas of \cite{Marino:2002fk} apply. It
  can be shown however that the {\it only} Seifert spaces for which the group
  of sheet transitions is finite, and whose planar resolvent thus gives rise to {\it algebraic}
  spectral curves, are precisely the Clifford--Klein 3-manifolds: there is no hope to
  extract an algebraic setup for the large $N$/B-model curve, even for the
  restriction to the trivial flat connection, for parabolic and hyperbolic
  Seifert 3-manifolds. In the elliptic case, instead, $\re^{W_{0,1}}$ is a root of an
  algebraic equation in $\bbC^\star \times \bbC^\star$:
\beq
P^\Gamma(x, \re^{W_{0,1}};t)=0
\eeq
where $t$ is the 't~Hooft parameter $g_s N$. It should be noticed
that this symmetrisation is not unique; however the freedom of choice here is
in bijection with the freedom of choice of an irreducible $\cG_\Gamma$-module
on the Toda side: in either case (see \cite{Borot:2015fxa,Brini:2017gfi} and \cite{MR1401779} respectively) it can
  be shown that these choices do not affect the calculation of the partition
  function, so we will henceforth drop the subscript $\rho$ from \cref{eq:Bcurve}.
\item For these cases, a computational tour-de-force leads to determine in
  full detail the spectral curve for
  type A \cite{Brini:2011wi}, D \cite{Borot:2014kda}, and $E_6$
  \cite{Borot:2015fxa}; substantial information can be extracted for
  $E_{7,8}$, with a full solution available in all cases in the limit $t\to
  0$. 
\item Now, proving the B-model side of the GOV correspondence for the LMO
  invariant amounts to finding a restriction $u_i(t)$ of the Toda hamiltonians of the
  ADE relativistic Toda chain such that 
\beq
\mathfrak{p}^\Gamma\l(x;
u_i+\delta_{i,\mathsf{k}}(y+u_0y^{-1}\r)|_{u_i=u_i(t)}=P^\Gamma(x, y;t)
\eeq
A detailed analysis of both spectral curves setup shows\footnote{Technically,
  the polynomials almost never match on the nose, but it can always be shown
  that upon restriction the Chern--Simons spectral curve arises as a
  non-trivial reducible component of a degenerate limit of the Toda curve.} that this is indeed
the case for all ADE types \cite{Borot:2015fxa,Brini:2017gfi}. 
\item Finally, having a matrix integral representation for the partition
  function of Chern--Simons theory on these spaces allows to rigorously derive
  the topological recursion of \cref{eq:toprecW,eq:toprecF} for the cumulants
  of the resulting distribution. On the other hand, on the B-model side, the
  topological recursion can be either regarded as the definition of the higher
  genus open/closed topological B-string on a family of curves, or, from a
  physics standpoint, it could be derived from the chiral boson theory on the
  spectral curve obtained from dimensional reduction of the BCOV Kodaira--Spencer
  theory of gravity \cite{Bershadsky:1993cx,Dijkgraaf:2007sx}. Since the two
  theories boil down to the same recursion with the same input datum, we reach
  the conclusion that they
  give rise to the same invariants to {\it all orders in $1/N$} and arbitrary
  colourings of the invariants. This yields an all-genus proof of the B-model
  side of the GOV correspondence for this type of manifolds, restricted to the
  LMO invariant.
\een

\label{sec:Bmod}

\section{Conclusions}

We have reviewed how a strict generalisation of the GOV correspondence that
bears all the ingredients of the simplest original setting of
\cite{Gopakumar:1998ki,Ooguri:1999bv}, including a geometric A-model theory
of open/closed Gromov--Witten invariants and an all-genus B-model theory
governed by the topological recursion on a specific spectral curve
setup, can be given for the case of Clifford--Klein 3-manifolds -- and these
alone, according to our remarks in the previous section. This opens severals
directions for future research, including an extension to the setting of
refined/categorified invariants (particularly on the B-model side), quantum
integrability, and the relation to gauge theories. We single out here three
more topics in particular on which we hope to report in the near future.

\subsubsection*{B-model general flat backgrounds}

Since most of the analysis of the previous section was restricted to the study
of the LMO invariant, it is natural to ask how to extend the GOV
strings correspondence to a general Chern--Simons vacuum, thus completing the
proof of the B-model version of the GOV correspondence. On the B-model side,
the family of relativistic Toda spectral curves was constructed in
\cite{Borot:2015fxa} for type $\mathrm{ADE}_{6,7}$, and the missing $\mathrm{E}_8$ case 
has recently been treated in detail in \cite{Brini:2017gfi}. The more difficult bit here is to provide a complete
large $N$ analysis of the matrix model, although this might be possible
by suitably rewriting the finite sums expressions of
\cite{Marino:2002fk,Blau:2013oha} in terms of ordinary eigenvalue models on
the real line. Establishing an explicit solution of the loop
equations for this matrix model in terms of the topological recursion applied
on the corresponding Toda spectral curve would give a full proof of the
B-side of the GOV correspondence. This would also shed light on some of the
difficulties encountereed in the analysis of arbitrary flat backgrounds to the
case of general lens spaces and non-SU(2) abelian quotients of the resolved
conifold \cite{iolpq}.

\subsubsection*{Non-toric ``remodeling-the-B-model''}
Another line of developement consists of extending the remodeling proposal of
\cite{Bouchard:2007ys} to the non-toric setting at hand for type D and
E. A first step here would be to fully spell out the computation of the disk
functions as in \cite{MR2861610,Brini:2011ij} for the case at hand, and then
derive the topological recursion from the analysis of the descendent theory.
A promising route would be to derive the $J$- and $R$-calibrations for the quantum cohomology of $\widehat{X}^\Gamma$ from the steepest descent analysis
of oscillating integrals of the Toda differential, as in \cite{Brini:2013zsa}, and then retrieve the
topological recursion from Givental's $R$-action on the associated
cohomological field theory \cite{DuninBarkowski:2012bw,Fang:2016svw}. This would lead to a
proof of the remodeling conjecture on an important class of examples, beyond the toric case.

\subsubsection*{The quantised McKay correspondence}

One of the more intriguing consequences of the B-model GOV correspondence is that several limits of
the B-model geometries of \cref{sec:Bmod} would provide a unified construction
of spectral curves relevant for both non-toric mirror symmetry and the theory
of Frobenius manifolds. In particular, the $u_0\to 0$ limit, corresponding to
the limit where the K\"ahler volume of the base $\bbP^1$ in
$\widehat{X}^\Gamma$ is sent to infinity, gives rise to a family of
1-dimensional Landau--Ginzburg models for the stack $[\bbC^2/\Gamma]$
and its crepant resolution: this would finally grant access to a host of
explicit computations on the descendent theory, which are likely to be
instrumental in the proof of the quantum McKay correspondence in full
generality \cite{MR2234886,MR2483931,MR2529944}. This would notably include the higher genus theory, by following
the arguments employed for the type A case in
\cite{Brini:2013zsa}. Moreover, Dubrovin's almost duality would relate these
mirrors to the LG formulation of the Frobenius manifold structure on the orbit
spaces of extended affine Weyl groups and ordinary Weyl groups associated to
simply-laced root systems upon considering, respectively, the relativistic and
non-relativistic limit of the Toda curves of \cref{sec:Bmod} \cite{Brini:2017gfi}. We plan to
further explore this in future work.

\bibliography{miabiblio.bib}
\end{document}